\documentstyle[aps,prd,preprint]{revtex} 
\begin{document} 
\draft
\preprint{imsc/98/06/30}
\title{Systematic technique for computing infrared properties of 2+1- and 3+1- 
dimensional Yang-Mills theory}
\author{ H.S.Sharatchandra \thanks{e-mail:sharat@imsc.ernet.in}}
\address{Institute of Mathematical Sciences,C.I.T campus, Taramani, 
Chennai 600-113.} 
\maketitle
\begin{abstract} 
New collective coordinates, related to the field at the `center'
of the monopoles, are proposed. A systematic computation of the infrared
properties of 2+1- and 3+1- dimensional Yang-Mills theory is now possible and
is related to solutions of classical equations with constraints at isolated
points. For 2+1-dimensional Yang-Mills theory, monopoles of a specific size
proportional to $g^{-2}$ dominate and a semiclassical technique is applicable. For
3+1-dimensional Yang-Mills theory, the formalism incorporates a monopole 
condensate naturally, and is therefore a correct starting point for
computations of confinement properties.
The method also provides a precise way of going beyond the dilute gas
approximation for instantons in 3+1-dimensions. Another algorithm, which 
uses only the quadratic form of the action and corrections via renormalized
perturbation theory, is proposed as a viable scheme of computation for all
length scales.
\end{abstract}
\pacs{PACS No.(s) 11.15-q, 11.15Tk}
\section{Introduction}
The ultraviolet behaviour of quantum chroodynamics are computable, thanks 
to asymptotic freedom. As a result QCD
can be confronted with experiments and has given us the confidence that it is the
correct theory of strong interactions. The theory is expected to become strong in the
infrared regime and lead to quark confinement. Strong coupling expansion of lattice
gauge theory \cite{wil} gives a strong reason to believe that linear confinement does take place.
It has been conjectured that the physical mechanism underlying this phenomenon is dual
superconductivity. What is sorely missing at present is a technique for realising this
and systematically computing the infrared properties of non-Abelian gauge
theories.Monte-Carlo simulations of lattice gauge theories have provided a valuable tool
for this purpose. However computing the numbers reliably for the the continuum 
limit is still difficult. An analytical technique is
always welcome.

Valuable hints to confining mechanism has come from the study of compact $U(1)$
lattice gauge theory \cite{p1,p2,b}.In the 2+1-dimensional case, monopole - 
anti-monopole plasma
is responsible for linear confinement as a consequence of Debye screening.In a seminal
work, Polyakov \cite{p2} has developed a semi-classical technique for computing the infrared
properties of 2+1-dimensional (continuum) Georgi-Glashow model. This is possible
because there is a finite energy stable classical solution ( the t'Hooft-Polyakov
monopole) of the Euclidean action. In principle the quantum corrections can be computed
to any order and are formally small for a range of parameters. Therefore we have
in principle an algorithm for computing the infrared properties.

Though it is generally felt that a similar mechanism is in operation for 2+1-dimensional Yang-Mills
theory, it is not clear how the ideas can be implemented. There is no finite
energy stable monopole solution and a semiclassical expansion scheme is not
evident.

The situation, as regards 3+1-dimensions,  is even worse. Compact U(1) lattice
gauge theory is equivalent to a system of (dual )photons \cite{p1} interacting with
monopole loops. This is made explicit by duality transformation \cite{b}. There is sufficient
reason to expect a condensate of monopoles \cite{b,gu} for a strong enough
coupling.
Again it is not clear as to how these ideas can be implemented for 3+1-dimensional
pure Yang-Mills theory.t'Hooft \cite{h1} has suggested that the monopoles of Yang-Mills theory
may be characterized by the zeroes of composite Higgs in the adjoint representation.
He has also advocated the idea of abelian projection for handling the
confinement properties: choose the unitary gauge for the
composite Higgs. Then the relevant degrees of freedom are the gauge bosons of the
maximal abelian subgroup and the monopoles. Unfortunately this has not yielded a viable
computational technique.

In this paper we propose a systematic scheme for computing infrared properties of 
2+1- and 3+1- dimensional Yang-Mills theory. In sec. 2 we discuss the finite
energy monopoles of Yang-Mills theory (without Higgs) and emphasize their 
topological nature which make them prime candidates for generating confinement.
The problem in pure gauge theory is that the energy of these monopoles
can be arbitrarily small (and not infinite as frequently made out to be).
Therefore the action favors a dense overlapping large size monopoles.
Conventional semi-classical technique, depending on finite energy stable
classical solution, is not applicable. We introduce  new collective coordinates 
(Sec 3). These are related to the behaviour of the gauge potential at the
`center' of the (anti-)monopole. These new collective coordinates make it possible 
to have a non-trivial minimum and a systematic semi-classical approximation, 
even though
there is no stable classical solution to the Euclidean action. What is relevant
now is the solution of the classical equations of motion with constraints at
{\sl isolated points} representing the `centers' of the (anti-)monopoles. 
Our method is closely related to the constrained  instanton
technique \cite{a}. But the constraint is realized in a much simpler manner, 
which makes computations much simpler.
 
Our collective coordinates are valuable in the traditional cases too (sec. 5): 
instanton gas in 4-(Euclidean) dimensional Yang-Mills theory \cite{dhn}, 
1+1-dimensional Heisenberg model \cite{bp}, or 1+1- dimensional Abelian Higgs 
model \cite{a}. The problem for handling instanton -anti-instanton
gas is now shifted to solving classical equation of motion with boundary constraints
specified at {\sl isolated points} corresponding to the `centers' of the instantons. 
As a consequence our technique is not limited by the dilute gas approximation.
The infrared divergence due to large size instantons 
encountered in the dilute gas approximation may be cured by the present technique.

In 2+1-dimensions, when the collective coordinates are small compared to the 
distance of seperation
between the monopoles, the relevant minimum of the action is given by \cite{g}
the Wu-Yang 
solution \cite{wy} outside a core of the size of the order of the collective coordinate. 
The core provides a form factor and gives a finite contribution to the action. Thus the
technique of Polyakov for 2+1-dimensional Georgi-Glashow model \cite{p2} is valid
{\sl in toto}, for this regime of collective coordinates, for the Yang-Mills case also. 
This explicitly justifies the proposal by Banks, Kogut and Myerson \cite{b} to use Wu-Yang solution with a form
factor. However this picture is no longer valid for large collective
coordinates.
Thankfully however, the competition between energy and entropy makes a particular 
size proportional to $g^{-2}$  of (anti-)monopoles contribute maximally to the 
functional integral (Sec. 4). Therefore it is possible to mimic the semiclassical calculation even though
there is no classically stable solution. Das and Wadia \cite{d} have proposed
that the many body effects may stabilize the size of the core. With our new
collective coordinates, the stability arises at the  semi-classical level itself,
and many body effectare not required.

The technique can be formally applied to 3+1-dimensional case also (sec. 6). Now
the `centers' of the monopoles trace out a world line. We have to sum over all 
such strings.In addition we have to sum over the collective coordinates 
of the monopoles all along the world line.

There are some crucial differences with the 2+1-dimensional case. We can have
monopoles  running all the way from infinite past and into infinite future.
However this can contribute to the functional integral only if they have
infinite size and zero energy for all but a finite duration of time. A
semi-classical technique as in 2+1-dimensions is not applicable.
On the other hand such configurations contributing to the functional integral 
naturally incorporate a condensate of monopoles, as is required for 
confinement. Thus our formulation appears to be the right starting point for the
confining aspects of non-Abelian gauge theories.

In sec.7, we point out that the procedure can be carried out with only the quadratic 
part of the action. We need solutions of linear equations with the appropriate 
boundary conditions. This may provide a viable scheme of computation for the 
confining aspects of non-Abelian gauge theories.

\section{Finite energy monopoles of Yang-Mills theory and  their `centers'}
Consider the Euclidean functional integral of 2+1 dimensional Yang-Mills theory,
\begin{equation}
Z=\int {\cal D}A_{i}^{a}(x)\:exp\left(-\frac{1}{2 g^{2}}\int
d^{3}x B_{i}^{a}(x) B_{i}^{a}(x)\right)
\label{3d}
\end{equation}
where $\{A_{i}^{a}(x),\;\;(i,a\:=1,2,3)\}$ is the Yang-Mills potential and
\begin{equation}\label{b}
B_{i}^{a}=\frac{1}{2}\epsilon_{ijk}(\partial_{j}A_{k}^{a}-
\partial_{k}A_{j}^{a}+\epsilon^{abc} A_{j}^{b} A_{k}^{c})
\end{equation}
is the field strength. In this case the coupling constant $g$ has the dimension
$1/2$ in mass. Perturbation expansion is obtained by a rescaling, $A
\rightarrow \sqrt g A$ and making an expansion in $g$. This gives massless
gluons. The theory is superrenormalizable as regards the ultraviolet
divergences. There are severe infrared divergences \cite{h2}. It is hoped that
this is indicative of non-perturbative effects leading to confinement.

As in the paradigm case of the 2-dimensional $xy$-model \cite{xy}, we may expect that
there are contributions to the functional integral from topologically
non-trivial configurations, which disorder the spin wave phase.This is
explicitly realised in the 2+1-dimensional Georgi-Glashow model \cite{p2}. Here there is
a finite energy stable monopole solution to the classical (Euclidean) equations of motion
whose effects can be computed by a semiclassical approximation and systematic
corrections to it. In the Bogomolni-Prasad-Sommerfield limit the solutions have
the explicit form
\begin{eqnarray}
A_i^a(x)= \epsilon_{iaj} \frac{x^j}{r^2}(1-\frac{vr}{sinh (vr)})  \nonumber \\
\phi^a(x)=\frac{x^a}{r^2} ((vr) coth(vr)-1) 
\label{m}
\end{eqnarray} 
where $v$ is the expectation value of the Higgs field (and $r=\sqrt {x^2}$).
$v^{-1}$ charecterizes the size of the monopole.
Non-trivial topological nature of the solution is characterized by the first
Chern class. $(2 \pi)^{-1} \int dS^i \phi^a(x) B_i^a(x) = 1$ for this solution.
This characterization uses
the asymptotic behaviour of the fields. There is an alternative way \cite{ar} which
uses the behaviour of the fields at the `center' of the 
monopole, given by the
isolated zeroes of $\phi$. The topological character is now provided by the
Poincare-Hopf index. $\phi^a = \rho^{-2} (x-x_0)^a + O(x-x_0)^2$ at the center 
$x = x_0$.

Consider only the gauge potential part of the above solution. This 
configuration gives a finite contribution to the action for any $v$.A 
simple way of seeing this is to note that energy $\int d^3x (( D_i \phi)^2 + B_i^2)$
is finite and $D_i \phi = B[A]_i$ for the Bogomolni limit.Therefore such 
configurations and fluctuations around it have to be included in the computation
of the functional integral and their effects investigated.But there are two 
major hurdles in doing this for Yang-mills theory:

i. Does the gauge potential configuration Eqn. \ref{m} by itself have a non-trivial topological content, so that it
may have a crucial effect on the massless gluons as in the 2-dimensional xy
model?

ii.Scaling implies that the only stable configuration of this kind corresponds
to $v=0$,  i.e. infinite size and zero energy. Therefore a semi-classical
method is not evidently applicable and it is not clear how they can be  handled.

We consider each of these issues in detail.

To characterize the monopoles of Yang-Mills theory, t'Hooft \cite{h3} has advocated 
the use of a scalar composite of gluons in the adjoint representation of the 
gauge group in place of $\phi^a$. We \cite{am} have proposed
an alternative characterization which uses the gauge potential directly.
Consider the gauge invariant composite $I_{ij}(x)=B_i^a B_j^a(x)$ which is a 
symmetric matrix at each $x$.  Monopoles are identified with $\sl isolated $ points where this matrix is
triply degenerate, $I_{ij}(x) = \rho^{-2} \delta_{ij}, x=x_p$. Consider the three orthonormal eigenfunctions
$\chi^A(x), A=1,2,3$. One of these eigenfunctions, say $A=1$,  will have a 
`radial'  behaviour \cite{s}, $\chi_i^1(x) = \rho^{-2} O_{ij} x^j + O(x^2)$ (where $O_{ij}$ 
is an arbitrary  orthogonal matrix)  as for the Higgs at the 'center' of a monopole.

The behaviour for the gauge potential at the `center' $x_0$ of the
monopoles which has the above properties is 
$A_i^a(x)= \rho^{-2} \epsilon_{iaj} (x-x_0)^j+O(x-x_0)^2$. Here $R$
is an parameter of dimension of length. In the Prasad-Sommerfield solution,
$\rho=\sqrt 6/v$ and reflects the size of the monopole. 
In particular the gauge field potential is non-singular at the
center $\rho$ is arbitrary and has no role to play in the
topological property of the monopole. All the topological content is coming
from the way the isospin and space indices are intricately mixed in the above
expression.

The topological character of these centers are also described by the effect on
Wilson loops \cite{go}. Consider any surface enclosing the center. Consider a class of
Wilson loops all starting at a fixed point on this surface and spanning the
surface. Being trace of an SU(2) matrix, the Wilson loop has the value $2  cos \theta$. The
angle $\theta$ is the magnetic flux. As the loops span the
surface the angle $\theta$,  regarded as a continuous function,  changes from 
zero to
$2 \pi$. {\sl This is true, however small the surface is, so long as it encloses the
center. The asymptotic behaviour of the monopole solution is not necessary for
this behaviour.} This also illustrates why such centers could play a crucial role
in confinement. A shift in the center can change the Wilson loop by a large
amount.

The magnetic field has the behaviour 
$B_i^a(x)=\rho^{-2} \delta_{ia}+O(x-x_0)$.
Thus $\rho$  characterizes the magnetic field strength at the center of the
monopole. Note also that the non-Abelian part 
$A \times A$ plays no role in this
behaviour. Thus the topological character is retained in perturbation
theory also. This will be used in a crucial way in Sec 7.

\section{New collective coordinates}
The contribution of the gauge potential in Eqn. \ref{m} to the action is 
proportional to $v/g^{-2}$. This leads to two problems. The stable configuration
corresponds to $v \rightarrow \infty$, i.e. infinite size, which is energetically
indistinguishable from the perturbative vacuum.Therefore, it a semi-classical 
technique is not evident. Also, large size monopoles have a very small action. 
Hence the problem in non-Abelian gauge theories without Higgs phenomenon is that 
one has to contend with very 
dense, very large sized overlapping  monopoles. This is in drastic contrast
to the corresponding Georgi-Glashow model \cite{p2} where the stable monopoles 
have large
actions and are therefore rare. We have to devise a technique which is not
tied to a dilute gas approximation.

We now propose new collective coordinates for handling the situation. Given
any arbitrary configuration of the gauge potential $A_i^a$, 
we may locate the isolated points where $I_{ij}(x_p)$ is triply degenerate. 
This gives the  centers  $x_{+p}, p=1,2, \ldots m$ of monopoles and,
$x_{-p}, p=1,2, \ldots n$ of anti-monopoles.
Summing over all configurations of the gauge potentials can be split into two
steps. We first consider configurations with specified locations $x_p^i$, the 
parameters $\rho_p$, and $O_p$ corresponding to the centers of monopoles and
anti-monopoles. We sum over all gauge potentials with these constraints. Then
we sum over all possible locations, size parameters and isospin orientations.
Thus the functional integral may be rewritten as
\begin{eqnarray}
Z=\sum_{m,n =0,1,2,\ldots} \Pi_{p=1}^m \Pi_{q=1}^n V_{m+n}
\int d^3x_{+p} \int_0^{\infty} \frac{d \rho_{+p}}{\rho_{+p}^4} \int dO_{+p}
\int d^3x_{-q} \int_0^{\infty} \frac{d\rho_{-q}}{\rho_{-q}^4}\int dO_{-q} \\ \nonumber
\int {\cal D}A'\:exp\left(-\frac{1}{2 g^{2}}\int
d^{3}x B_{i}^{a}(x) B_{i}^{a}(x)\right)
\label{n}
\end{eqnarray}
\noindent
$DA'$ means integration over all gauge field configurations with the variables
$x,\rho,O$ held fixed. $V_{m+n}$ is a numerical constant.
The factor $\rho^{-4}$ in the measure is as in the instanton gas, except that now
we are in 2+1-dimensions. It comes from a dimensional analysis \cite{h4,dhn}
and can be justified independently of the semi-classical calculation \cite{s}.
It turns out to be crucial for calculations in 2+1-dimensions (Sec 4).

This way of summing over the configurations have many advantages.
It goes beyond the perturbation theory because the effects of monopoles are
explicitly taken into account. It is also possible to do a semi-classical
approximation now. This would not have been possible if hadn't seperated out the
collective coordinate $\rho$. The extremum of the action would be zero,
corresponding to a monopole of infinite size. It is not differentiated from
the perturbative vaccum. On the other hand  we may now hold
the position and the size parameter fixed. Therefore  the minimum of the action
corresponds to the solution of the classical Yang-Mills equations with these
boundary conditions. In Ref. \cite{g} it is shown that this gives a solution
which is exactly the Wu-Yang solution \cite{wy} outside a core of size $O(\rho)$. 
Inside, it gives a form factor because of which the solution is of a finite
energy, in contrast to the Wu-Yang solution. The gauge potential has a
discontinuous derivative at the core. This is a case where the extremum is
realized on the boundary of the space of continuous functions. The
discontinuity is innocuous. The action is finite and the effects of
fluctuations can be systematically computed. (It is analogous to the
Schrodinger equation in a square well potential.)

Our technique is closely related to the constraint instanton formalism
\cite{a}. However we realize the constraints in a much simpler fashion.
The constraint is at isolated points. and we have to simply solve classical
equations with such boundary conditions. This makes the present proposal a
viable computational scheme. 

Banks, Kogut and Myerson \cite{b} have considered Wu-Yang solutions with a core
as possible candidates for disordering the spin wave phase in 2+1-dimensional
Yang-Mills theory.Our collective coordinates realize this in a precise manner.
The core is obtained naturally by the extremization. ( In Ref. \cite{b},
smoothness is demanded at the boundary of the core. But this forces the
solution to have a singularity at the center of the monopole \cite{g}.)

We may also consider extremization in case of many monopoles and
anti-monopoles. If the size parameters $\rho$ are all rather less than the 
distance of seperation between the monopoles, then the extremum is again
similar to the one monopole case: Outside the cores of size $\rho$, the solution
is a non-Abelian gauge rotation of the solution in the corresponding Maxwell 
theory \cite{g}. This corresponds exactly to the dilute monopole approximation 
of Polyakov \cite{p2}. Thus Polyakov's computation can be
taken over {\sl in toto} for this range of collective coordinates.

The most important advantage of our collective coordinates is that it provides
a precise way of going beyond a dilute gas approximation.When the size
parameters are large, we no longer get the Wu-Yang solution with a core.
However the problem of extremization is still well posed and the extremum is
well defined. This accounts for interactions in a precise way and we are not
bound by a Coulomb gas approximation.

\section{Mean size of the monopoles and a semi-classical technique}
For a fixed $\rho$, the minimum of the action is \cite{g}  
$\kappa/ (g^2 \rho)$ where $\kappa  \sim 2.3$. This favors very large size
monopoles which have very small energies. However the scale factor $\rho^{-4}$
in the measure  favors small $\rho$. Due to a competition between these `energy'
and `entropy'  factors, a mean size of monopoles dominate. This size is
$g \rho_0 =\kappa/2 $. This makes it possible to have a computation in exact 
analogy with the semiclassical calculation of Polyakov \cite{p2} for the
Georgi-Glashow model, even though the only stable solution is of 
zero energy. We need to only include the contribution to fluctuation in the 
size as a correction.

Das and Wadia \cite{d} have proposed that quantum effects may stabilize the 
size of monopoles (with a core) for 2+1-dimensional Yang-Mills theory. In
our technique here, the stability comes about in the semi-classical limit
itself, as a result of a competition between energy and entropy.

\section{3+1-dimensional Yang-Mills theory: monopole condensate}
We now consider 3+1-dimensional Yang-Mills theory.At each instant of time the
configuration of the Yang-Mills potential $A_i^a(x)$ can be characterized by
the location of the monopoles and the parameters $\rho$ and $O$.
Because of the topological character, the monopole `centers'
have to continue in time. The only way they can disappear is by annihilation of
a monopole with an anti-monopole. Thus the four-dimensional configuration of
the Yang-Mills potentials can be characterized by the strings $x^{\mu}(\tau)$
representing the `centers' of (anti-)monopoles and the remaining collective
coordinates $\rho(\tau), O(\tau)$.
These strings can be either closed, representing monopole-antimonopole loops
or they can be open ended, representing (anti-)monopoles coming from the infinite
past and going off indefinitely into the future.We have to sum over all such strings 
and along each string,  all values of $\rho(\tau), O(\tau)$, in order to compute
the full functional integral.Thus the functional integral can be explicitly
written as,
\begin{eqnarray}
Z=\sum_{m,n =0,1,2,\ldots} \Pi_{p=1}^m \Pi_{q=1}^n 
\int Dx_{+p}(\tau) \int \frac{D\rho_{+p}(\tau)} {\rho_{+p}(\tau)^4}
\int DO_{+p}(\tau)
\int Dx_{-q}(\tau) \frac{D\rho_{-q}(\tau)} {\rho_{-q}(\tau)^4}
\int DO_{-q}(\tau)  \\ \nonumber
\int\:{\cal D}
A_{i}^{a}\;exp(-\frac{1}{4g^{2}}\int d^4x \:F_{ij}^a(x) F{ij}^a)
\label{4d}
\end{eqnarray}
\noindent

Consider an
open ended string. It has an infinite action and hence irrelevant if the
parameter $\rho$ is finite at large times. Only configurations which have 
$\rho(\tau)=\infty$ (i.e., infinite size and zero energy) for 
most of the history except for a finite interval of time, are relevant. We 
propose that such open-ended strings are the `instantons' relevant for confinement. 

Note that the competition between entropy and energy does not favor a mean
size now (because a non-zero mean size gives infinite action), in contrast to the 
2+1-dimensional case. Now the relevant
configurations are monopoles which are of infinite size and zero energy for
most of the time and which have brief fluctuations to a finite size.This is an important
difference, and changes the way to tackle the problem.

We now argue that this representation of the functional integral suggests a
condensate of monopoles, as required for quark confinement. Introduce
creation and annihilation operators $\phi^*(x, \theta), \phi(x, \theta)$ for the
monopoles. Here $\theta$ labels the isospin orientation of the monopole,
$O=exp(i \theta^a T^a /2)$. The operator $\phi(x, \theta)$ in the functional 
integral has the effect of summing over only those configurations where a
monopole string with isopin orientation $\theta$ ends at $x$ or an anti-monopole
string with isospin orientation $ - \theta$ begins at $x$. Note that 
open-ended monopole and anti-monopole strings can occur independently.
In our way of computing 
the functional integral, the expectation value $ \langle \phi(x, \theta) \rangle$
is finite, because open-ended strings with occasional finite size contribute
a non-zero action, and have to be included in the calculation. 
Thus a condensate of monopoles is present.
Therefore we may expect this to provide a correct starting point for
computations of the confining properties.

Efficient ways for summing over the strings and the collective coordinates are
yet to be devised.

\section{Beyond the dilute instanton gas approximation}
Our techniques can be applied fruitfully to the usual cases also: for
example, instanton gas in 3+1-dimensional Yang-Mills theory \cite{dhn}, 
SO(3) Heisenberg model \cite{bp} or even the 1+1-dimensional Abelian Higgs model.
Consider the BPST instantons. As in sec 2 for the monopoles, the 
topology can be located at the `centers' of
the instantons, $A_{\mu}^a(x)= \rho_p^{-2}\eta_{a \mu \nu} (x-x_p)^{\nu}+O(x-x_p)^2$.
Now it is possible to go beyond the dilute gas approximation. We have to solve
Yang-Mills equations with the above boundary conditions. The major problem 
with the dilute gas approximation in instanton gas is that the large size
instantons have a diverging contribution. It is precisely for the large sized
overlapping instantons that the dilute instanton gas approximation is not to
be trusted. Our technique is free of this problem. It is to be seen whether the
infrared divergencies are avoided in the present approach.

\section{Non-perturbative effects from the quadratic part of the action}
In our approach, the problem of overlapping dense monopoles (instantons) is shifted to the
solution of classical equations of motion with constraints at isolated points
representing the centers of the monopoles (instantons). Solving non-linear equations 
with constraints at multiple points is not easy.
We now propose an algorithm, which starts with 
only the quadratic part of the action, for computing the effects
of monopoles (instantons). The quantum corrections can then be systematically included
using renormalized perturbation expansion. This may provide a
systematic technique which accounts for both asymptotic freedom and confining aspects.

We have characterized the `centers' of the monopoles by the points where $B_i^a(x) B_j^a(x)$
is triply degenerate. The behaviour of the gauge potential Eqn. \ref{m}, is such that
the quadratic term $A \times A$ of $B$ is irrelevant for this property. Also, this
boundary condition is consistent with a finite value for the
quadratic part of the action. Thus we may start with only the quadratic part of
the action and sum over configurations with fixed locations and collective 
coordinates $\rho, O$ of monopoles. We may use renormalized perturbation theory in $g$ to
calculate corrections to this. This requires simply the solutions of 
Maxwell's equations with boundary conditions Eqn. \ref{c}, and 
provides a viable calculation. ( As i the earlier case, we need solutions which 
have discontinuous derivatives.) We now argue that this may suffice for 
obtaining the confining properties. In Polyakov's semi-classical technique
\cite{p2}, area law for the Wilson loop arises as follows. The monopoles and
anti-monopoles of the plasma rearrange on either sides of the Wilson loop to
form a dipole sheet. As discussed in sec 3, as the monopole cuts through 
any sheet spanning the Wilson loop, the contribution changes by a large amount.
This is not just the property of the asymptotic Wu-Yang solution, but is
a property associated with the behaviour of the gauge potential at the `center'
of the monopole. Thus for confinement, it may suffice if the `centers' of
monopoles and anti-monopoles in the plasma correlate with the Wilson loop resulting in a
change in energy proportional to the area of the Wison loop.This is likely in the
above approach.

\section{Summary}
The problem in addressing confinement in non-Abelian gauge theories without 
Higgs fields is to identify the monopole configurations and to devise a technique for
summing over all sizes of monopoles. We have proposed a solution to both these problems.
We have emphasized the topology associated with the `centers' of 
monopoles and instantons.We have used this for computing the effects of topological 
objects in the functional integral. Our algorithm is applicable even when there are no
finite action stable classical solutions. Moreover it provides a viable method
for going beyond the dilute instanton approximation. Our way of accounting for
the monopoles provides a semi-classical technique for calculating the 
confining properties of 2+1-dimensional Yang-Mills theory. Wu-Yang monopoles 
having a core with a specific form factor dominate due to a competition between 
`energy' and `entropy'. In case of 3+1-dimensional Yang-Mills theory, we have 
proposed that occasional fluctuations into a finite size of monopoles which
are of infinite size and zero energy for most of the time are responsible for
confinement. We have argued that our way of handling the 
functional integral has a condensate of monopoles automatically. As a consequence, we expect
a computational tool for confinement. We have also argued for a much simpler
algorithm which starts with only the quadratic part of the action and uses
renormalized perturbation theory for obtaining both asymptotic freedom and
confining properties.
We believe that this provides a viable computational tool
for the quantum chromodynamic theory of strong interactions.


\end{document}